\documentclass[aps,pre,reprint,showpacs,showkeys, superscriptaddress,
floats,floatfix,preprintnumbers,longbibliography]{revtex4-2}
\usepackage{amssymb,amsmath}
\usepackage{bm,url,graphics,graphicx,subfigure,epsfig,rotating} 
\usepackage{mathrsfs} 
\usepackage{color,soul}
\graphicspath{{./}{Figs/}}
\usepackage{braket}
\usepackage{hyperref}
\usepackage{multirow}
\usepackage{makecell}
\usepackage[normalem]{ulem}   
\usepackage[dvipsnames, table, xcdraw]{xcolor}
\usepackage{relsize}

\def \Sh  {\mathcal{S}_{\rm h}}
\def \tauxx {\tau_{\bm{x}}}
\def \DD  {D_{\rm D}}
\def \DH  {D_{\rm H}}

\def \zbulk {\zeta_{\rm bulk}}
\def \deli {\partial_{i}}
\def \delj {\partial_{j}}
\def \delk {\partial_{k}}
\def \ui   {u_{i}}
\def \uj   {u_{j}}
\def \uk   {u_{k}}
\def \du {\delta u}

\def \LI {L_{\rm I}}

\def \zhat {\hat{z}}

\def \Sp  {S_{\rm p}}

\def \km   {k_{\rm m}}

\def \uu  {\bm{u}}
\def \kk  {{\bm k}}

\def \ff  {{\bm f}}

\def \oo  {{\bm \omega}}

\def  \xx  {{\bm x}}
\def \rr   {\bm{r}}

\def \XXm  {\bm{X}^{\rm m}}

\def \cs {c_{\rm s}}

\def \grad {{\bm \nabla}}
\def \curl {{\bm \nabla} \times}
\def \dive {{\bm \nabla}\cdot}

\newcommand{\avg}[1]{\left\langle #1\right\rangle}

\def \Rey  {\mbox{Re}}

\def \Ma  {\mbox{Ma}}

\def \kf  {k_{\rm f}}

\def \urms  {u_{\rm rms}}

\def \zetap {\zeta_{\rm p}}

\newcommand{\eq}[1]{~(\ref{#1})}

\newcommand{\Fig}[1]{Fig.~(\ref{#1})}
\newcommand{\subfig}[2]{Fig.~(\ref{#1}#2)}

\newcommand{\bfig}{\begin{figure}}
\newcommand{\efig}{\end{figure}}
\newcommand{\bc}{\begin{center}}
\newcommand{\ec}{\end{center}}
\newcommand{\bea}{\begin{eqnarray}}
\newcommand{\eea}{\end{eqnarray}}

\def \Vvec {{\bm V}}
\def \TE {T_{\rm E}}
\def \uvec {{\bm{u}}}
\def \xvec {{\bm{x}}}
\def \fvec {{\bm{f}}}
\def \kvec {{\bm{k}}}

\def \urms {u_{\rm rms}}

\def \zetap {\zeta_p}

\def \zetaps {\zeta_p^{\rm s}}
\def \zetapc {\zeta_p^{\rm c}}

\def \tauD {\tau_{\rm D}}
\def \tauH {\tau_{\rm H}}
\def \mD {\mathcal{D}}

\def \chip {\chi_{\rm{p}}}
\def \chiDp {\chi^{\rm D}_{\rm p}}
\def \chiHp {\chi^{\rm H}_{\rm p}}

\def \chiHp {\chi^{\rm H}_{\rm p}}
\def \chiDp {\chi^{\rm D}_{\rm p}}

\def \zetat {\zeta_{\rm 2}}

\def \dup   {\delta u_{\parallel}}
\def \Es {E_{\rm s}}
\def \Ec {E_{\rm c}}

\def \Spc {S^{\rm c}_{\rm p}}
\def \Sps {S^{\rm s}_{\rm p}}

\def \divu {\nabla\cdot{\bf u}}
\def \us {\bm{u}^{\rm s}}
\def \uc {\bm{u}^{\rm c}}
\def \Es {E_{\rm s}}
\def \Ec {E_{\rm c}}

\newcommand{\SMat}[1]{Supplementary Material \cite{SM}}

\begin{document}
\title{Intermittency and non-universality of pair dispersion in
  isothermal compressible turbulence}
\author{Sadhitro De}
\email{sadhitro.de@physics.ox.ac.uk}
\affiliation{Department of Physics, University of Oxford, Oxford,
  United Kingdom.}
\affiliation{Department of Physics, Indian Institute of Science,
  Bangalore 560012, India.}
\author{Dhrubaditya Mitra}
\email{dhrubaditya.mitra@su.se}
\affiliation{Nordita, KTH Royal Institute of Technology and
  Stockholm University, Hannes Alfv\'ens v\"ag 12, 10691 Stockholm, Sweden}
\author{Rahul Pandit}
\email{rahul@iisc.ac.in}
\affiliation{Department of Physics, Indian Institute of Science,
  Bangalore 560012, India.}
\date{\today}
\begin{abstract}
  We study  statistical properties of the pair dispersion of Lagrangian
  particles in transonic to supersonic compressible turbulence of an
  isothermal ideal gas in two dimensions, driven by large-scale solenoidal
  and irrotational stirring forces, via direct numerical simulations. 
  We find that the scaling exponents of the order-$p$ negative moments of
  the doubling and halving times of pair separations are nonlinear functions
  of $p$.
  Furthermore, the doubling and halving time statistics are different. 
  The halving-time exponents are universal -- they satisfy their multifractal
  model-based prediction, irrespective of the nature of the stirring. 
  The doubling-time exponents are not. 
  In the solenoidally-stirred flows, the doubling time exponents can be
  expressed solely in terms of the multifractal scaling exponents
  obtained from the structure functions of the  solenoidal component of the
  velocity. Moreover, they depend strongly on the Mach number, $\Ma$. 
  In contrast, in the irrotationally-stirred flows, the doubling-time exponents
  do not satisfy any known multifractal model-based relation, and are
  independent of $\Ma$.
  Our work is a generalization of Richardson's law of pair dispersion to
  compressible turbulence. 
\end{abstract}
\keywords{Compressible turbulence; Particle/Fluid Flow}
\maketitle
One of the pillars of our understanding of turbulence is 
 \textit{Richardson's law}~\citep{richardson1926atmospheric}:
$\avg{R^2(t)}\sim t^{3}$, where 
$R$ is the separation between a pair of Lagrangian particles or tracers,
$t$ is time, and $\avg{\cdot}$ denotes
averaging over the statistically stationary turbulent state.
This law holds for incompressible flows if $R$ lies within the
inertial range of length scales, across which the nonlinear interactions
dominate over viscous dissipation.
Here, we seek a generalization of Richardson's law to
compressible turbulent flows, which are prevalent
in many astrophysical
systems~\citep{Mordecai2004Review,McKee2007SF_Review,Elmegreen2004Review_ISM}. 

We begin by summarizing the present status of  Richardson's law in
incompressible turbulence. 
Theoretically, it can be understood as a modeling of the evolution of $R(t)$
by a diffusive process~\citep{richardson1926atmospheric, benzi2023lectures} with a diffusion
constant, $K(R) \sim R^{4/3}$, that stems~\citep{benzi2023lectures} from Kolmogorov's 1941
(K41) theory~\citep{K41}. 
Stated differently, Richardson's law implies that
there exists a dynamic scaling exponent, $z=2/3$,
such that typical time scales, $\tau(r)$, of turbulent eddies
of size $r$ scale as $\tau(r) \sim r^{z}$. 
It is now well established, from experiments and
direct numerical simulations (DNS), that
K41 must be generalized to include intermittency, i.e.,
the $p$-th moment of the Eulerian velocity difference, $\du$, across a
length scale $r$ (the order-$p$ velocity structure function), scales as 
$\avg{\delta u^p(r)}\sim r^{\zetap}$,
where $\zetap$ is a nonlinear concave function of $p$~\citep{frisch}. 
This also dictates that turbulence must, in fact, have an infinite number of
dynamic exponents~\citep{l1997temporal, MitraDyn, mitra2005dynamics,
  SSR_NJP, SSR_PRL, PanditDynRev}, suggesting the presence of intermittency
corrections to Richardson's law,
which itself enjoys feeble support from experiments~\citep{ott2000experimental,
 shnapp2018generalization, elsinga2022non, Tan2022PairDisp,
 shnapp2023universal} and DNS~\citep{salazar2009two,bitane2012time,
  bitane2013geometry, bragg2016forward, buaria2015characteristics,
  elsinga2022non} -- the observed scaling range of $\avg{R^2}$ is at most a
decade.
A better approach to investigate the intermittency of pair dispersion in
turbulence is to use \textit{exit times},
defined as the time, $\tau_{\sigma}(R)$, it takes for $R$ to cross
$\sigma R$ \textit{for the first time}, where $\sigma$ is fixed. 
The choice $\sigma>1$ corresponds to the
\textit{doubling times}\footnote{To be accurate,
  $\sigma=2$ means ``doubling''. Nevertheless, in practice, any $\sigma>1$ is
  used~\citep{Bofetta2002_RichardIntermitt}}, $\tauD$,
for which~\citep{Bofetta1999_Doubling,
  Bofetta2002_RichardIntermitt, PairDisp_Bofetta2002}
\begin{equation}
    \avg{\tauD^{-p}(R)} \sim R^{-\chiDp}\,,\quad\text{where}\quad 
    \chiDp=p-\zetap\,.
    \label{eq:tauD_Ch4}
\end{equation}
This bridge relation between the dynamic exponents, $\chiDp$, and
$\zetap$ has been obtained by using the multifractal model of
turbulence.  

In compressible flows, in addition to doubling times, we need to define
\textit{halving times} 
$\tauH(R)$, for which $\sigma<1$ and the corresponding scaling exponents are
$\chiHp$. 
As the multifractal model does not distinguish between the doubling and
halving times, $\chiHp$ are also expected naively to satisfy the
same bridge relation, \eq{eq:tauD_Ch4}. 
Although there have been many studies of the statistics of velocity increments
in compressible turbulence~\citep{Schmidt2008Intermit, Konstandin2012EulLag,
  Kritsuk2007Intermit2, wang2012scaling,wang2017scaling,
  jagannathan2016reynolds}
and tracer dispersion in compressible surface flows are known
not to obey Richardson's law~\citep{Cressman_2004_Surf_EPL}, 
the intermittency of tracer pair dispersion have not been studied.
We explore this via simulations of forced compressible turbulence of an
isothermal ideal gas, governed by the Navier-Stokes equations in
\textit{two dimensions}. 

In 2D compressible turbulence forced at an intermediate
  scale, Refs.~\citep{falkovich2017vortices, kritsuk2019energy} had obtained
  two cascades of the kinetic energy: an inverse cascade from the energy
  injection scale to large length scales and a forward cascade to small length scales.
  In what follows, we choose the energy-injection scales
  to be close to the linear size  of our simulation domain; and we focus on the
forward- cascade regime, which we show to be intermittent. 

\begin{figure*}[t]
    \centering
    \includegraphics[width=0.95\linewidth]{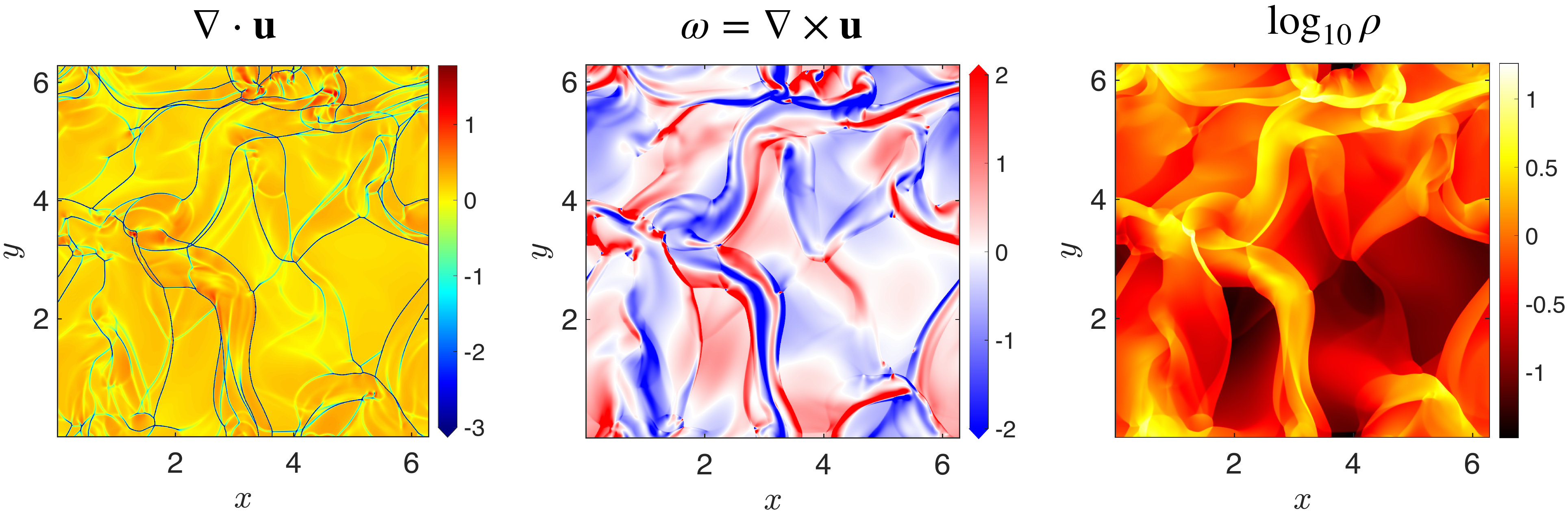}
    \caption{\small \textbf{Typical snapshots} of $\divu$,
      the vorticity $\oo=\curl\uu$, and $\log_{10}\rho$, for solenoidally-driven
      supersonic turbulence, run S2
      ($\dive\uu$ and $\oo$ are normalized by their respective
      root-mean-square values).
    Thin filaments in the $\dive\uu$ profile are the shocks.
    The vorticity, $\oo$, is large in the vicinity of curved shocks.
    There are strong density gradients across shocks.
    }
    \label{fig:snap}
\end{figure*}
We unveil a power-law growth of $\avg{R^2(t)}$,
but the scaling regime is too
  small to make robust quantitative measurements.
Thus, we resort to the exit-time approach to
calculate the dynamic exponents $\chiHp$ and $\chiDp$. 
We find that $\chiHp$ obey the multifractal model-based
bridge relation and are universal, i.e., does not depend
on how the turbulence in forced. 
But contrary to the prediction of the multifractal model
$\chiDp \neq \chiHp$.
Moreoever, $\chiDp$ is not universal. 
If the external force is solenoidal, then:
(a) $\chiDp = p - \zetaps$,
where $\zetaps$ are the scaling exponents of \textit{structure functions
  of the solenoidal component of the velocity field};
and
(b) $\chiDp$ depend on the Mach number ($\Ma$). 
If the force is purely irrotational, $\chiDp$
neither follows any known bridge relation, nor depends on $\Ma$.
We further identify, through the multifractal model,
the dimensions of the structures that play the
dominant role in pair dispersion. 

We perform pseudospectral DNS of the equations of
compressible, isothermal gas with a random white-in-time
external force (at the Fourier modes with
$\mid\kk\mid = \kf = 3$) on a 2D doubly-periodic square domain
of side $L=2\pi$ with $N^2=4096^2$ collocation points.
The assumption of isothermality holds when the gas is optically
thin (transparent to radiation) in which case the heat generated by the
dissipation at shocks is rapidly radiated away. This is generally valid
for molecular clouds which have low density and solar
metallicity~\citep{Mordecai2004Review}.
In addition to shear viscosity, we use a bulk viscosity
which is zero everywhere except near the
shocks~\citep{vonneumann1950method}.
Shocks are generally resolved across $4$ to $10$ grid points. 
After the flow reaches a statistically stationary state we seed it
uniformly with $n = N^2$ Lagrangian particles and track them as the
flow evolves. 
They cluster at the shocks and their number density
is proportional to $\rho(\xvec,t)$.

Using the Helmholtz decomposition, we write
$\uu =\us+\uc$, where $\us$ and $\uc$ are, respectively,
the solenoidal ($\nabla\cdot\us=0$) and irrotational ($\nabla\times\uc=0$)
components of the flow.
The relative importance of $\us$ and $\uc$ in the total
kinetic energy is measured by the solenoidal ratio,
$\Psi\equiv \avg{\rho|\us|^2}/\avg{\rho|\uvec|^2}$.
See Appendix~\ref{sec:numsim} for further details on the numerical
method. 

We perform four sets of DNSs,
S1 and S2 for solenoidal external force and
C1 and C2 for irrotational external
force, see Table~\ref{tab:para}
for all the parameters of the runs.
Typically, subsonic and supersonic compressible flow can be
qualitatively different.
Hence we perform a set of simulations that are transsonic, Runs
S1 and C1, $\Ma \simeq 1$, and another set 
S2 and C2 are  supersonic, $\Ma > 1$.
In \Fig{fig:snap} we show representative pseudocolour plots of
$\dive\uu$, $\oo = \curl\uu$, and $\log_{10}\rho$ for the run
S2; see \Fig{fig:Snap} for representative snapshots
The shocks are clearly visible as filamentary structures with large negative
values of $\dive\uu$.
Naively, we may expect that the presence of the shocks would not
show up in pseudocolour plot of $\omega$, but this not the
case because although a shock front that is a straight line does
not contribute to vorticity, \textit{curved} shock fronts do. 
In our simulations,
(a) the density gradients are larger for the supersonic cases
compared to the transonic ones, and
(b) the vorticity is more uniformly distributed in the transonic
cases than the supersonic ones.

We can define the $p$-th order longitudinal
\textit{equal-time} structures functions of
$\uu$ as well as its components, $\us$ and $\uc$:
\begin{subequations}
  \begin{align}
    &\delta u_{\parallel}(r) \equiv
               \left[\uu(\xx+\rr)-\uu(\xx)\right]\cdot\frac{\rr}{r}\/, \\
    &\text{and}\quad \delta u_{\parallel}^{\rm{s,c}}(r) \equiv
               \left[\uu^{\rm{s,c}}(\xx+\rr)-\uu^{\rm{s,c}}(\xx)\right]
               \cdot\frac{\rr}{r}\/; \label{eq:dup}\\
     &\text{with}\quad\Sp(r)=\avg{| \delta u(r)_{\parallel} |^p}\/,
     \quad \Sp^{\rm s,c}(r)=\avg{|\delta u(r)^{\rm{s,c}}_{\parallel} |^p}\/,
                  \label{eq:Sp_Ch4}\\
      &\text{and}\quad \Sp(r)\sim r^{\zetap}\/,\quad  \Sps(r)\sim r^{\zetaps}\/,
                          \quad \Spc(r)\sim r^{\zetapc}\/.
    \end{align}
    \label{eq:Spr}
\end{subequations}
The scaling relations hold for $\eta\ll r\ll \LI$, where $\eta$ is the energy
dissipation scale. 
All the \textit{equal-time} exponents, $\zetap$, $\zetaps$ and $\zetapc$ are
nonlinear monotone increasing functions of
$p$~\citep{Konstandin2012EulLag, wang2017scaling, wang2012scaling}.
The spectra of turbulent fluctuations and the equal-time structures
functions are shown in Appendices \ref{sec:spec} and \ref{sec:stfun}
respectively. 

We calculate $\avg{R^2(t)}$ by averaging, at every instant, over the
$2n$ separations where $n=N^2$ is the number of tracers.
See \Fig{fig:Rsquare}
for plots of $\avg{R^2(t)}$ versus time for all the runs. 
At early times, $\avg{R^2(t)}$ grows exponentially for all the
runs.
This range of exponential growth is very small for C1 and C2. 
Similar results~\citep{Jullien2003DipsersionExp} have been obtained for
incompressible 2D turbulence.
Exponential growth up to a length scale typically implies that the flow
across is smooth up to that scale. 
This holds for our simulations because at such small time scales most of the
pair separations lie away from shocks.
At late times the behaviour of $\avg{R^2}$ is
not universal, it depends on precisely how the DNS is forced.

If $R(t)$ lies in the inertial range, then
$R^2 \sim t^{2/(1-h)}$, where $h$ is the local H\"older exponent of the velocity
fluctuations in space. 
If we use Kolmogorov's simple scaling hypothesis, $\zetat=2h$ which yields
$\avg{R^2(t)} \sim t^{2z}$, where $z=1/(1-\zetat/2)$.
In incompressible flows $\zetat = 2/3$ and $z=3/2$
that corresponds to Richardson's law.
We calculate $\zetat$ from the scaling of the second order structure
functions, $S_{\rm 2}(r)$.
We find that there is a range of scales over which $\avg{R^2}\sim t^{2z}$ in
S1, S2 and C2; although not in C1. 
Crucially, in every case, the scaling regime is too small to make robust
  quantitative measurements of the values of $z$ from the $\avg{R^2(t)}$
  versus $t$ plots.
Hence,  we turn to exit-times. 

\begin{figure}
    \centering
    \includegraphics[width=0.48\linewidth]{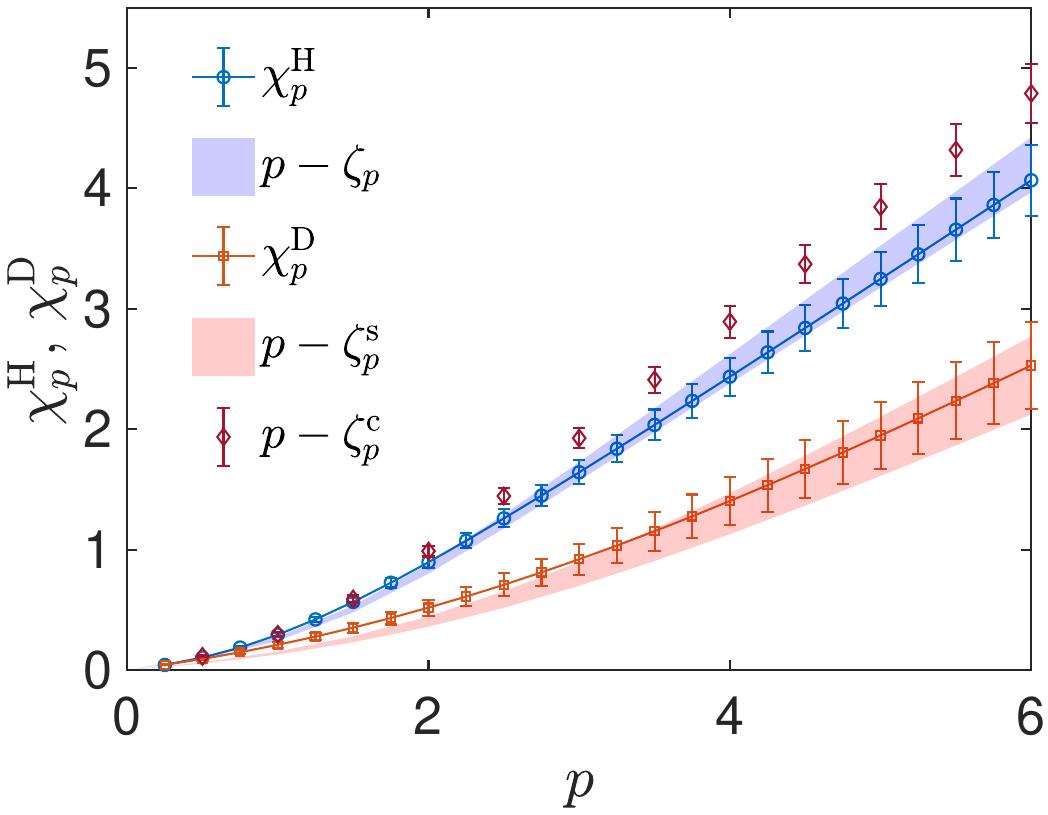}
    \includegraphics[width=0.48\linewidth]{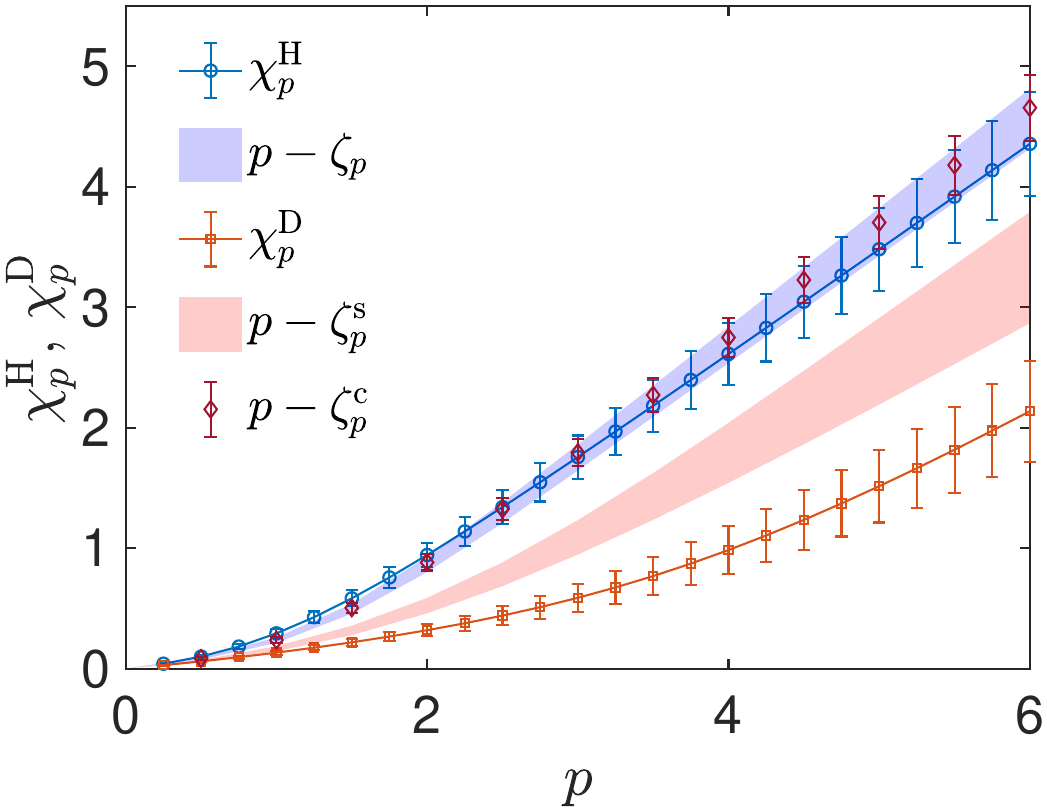}
    \caption{\small \textbf{Dynamic exponents} for doubling and halving times
      (see text) $\chiDp$ and $\chiHp$, respectively, as functions of the
      order $p$ 
      for the two supersonic runs [S2 (left panel) and C2 (right panel)]. 
      The blue shaded region corresponds to the bridge relation, $p-\zetap$,
      derived from the multifractal model.
      The pink shaded region corresponds to the expression $p-\zetaps$.
      In both S2 and C2, the multifractal prediction for $\chiHp$ is correct
      within errorbars, but not for $\chiDp$. For S2
      $\chiDp = p -\zetaps$ is correct within errorbars.
      For C2 $\chiDp$ does not satisfy any known bridge relation. 
    }
    \label{fig:Chi}
\end{figure}

In the stationary state, we choose $2n$ tracer pairs whose
relative distance $R$ lies in the inertial range and
then calculate $\tauH(R)$ and $\tauD(R)$ as the
first instant at which $R$ goes below $3R/4$ or above $3R/2$,
respectively~\footnote{To calculate $\avg{R^2(t)}$ we use initial
    values of $R$ to be approximately one grid spacing. But
    to calculate exit time we must start with $R$ within the
    inertial range}.
The statistics of exit--times is expected to show scaling (or multiscaling)
if the distances $R$, $3R/2$ and $3R/4$ lie
within the inertial
range~\footnote{As long as this condition is satisfied, we expect
  our results to be independent of the particular choice of the thresholds,
$3R/2$ or $3R/4$.}.
Specifically, the  inertial range is defined as the range of scales  over
which the equal--time structure functions $\Sp(r)$, $\Sps(r)$ and
$\Spc(r)$ [see \eq{eq:Spr}] show scaling.

We find that the negative moments of the exit-time distributions scale as
\begin{equation}
  \avg{\tauD^{-p}(R)}\sim R^{-\chiDp} \quad\text{and}\quad
  \avg{\tauH^{-p}(R)}\sim R^{-\chiHp}\,.
  \label{eq:tau}
\end{equation}
over almost a decade in $R$. 
We extract the exponents, $\chiDp$, $\chiHp$, $\zetap$, $\zetapc$ and $\zetaps$,
by linear least--square fitting of the log-log plots of $\avg{\tauD^{-p}(R)}$,
$\avg{\tauH^{-p}(R)}$, $\Sp(r)$, $\Sps(r)$ and $\Spc(r)$, respectively. 
In \Fig{fig:Chi} we plot the dynamic exponents, $\chiDp$ and $\chiHp$,
for the runs S2 and C2, up to $p=6$. 
Both are nonlinear functions of $p$, indicating \textit{dynamic multiscaling}.
The spreads of the shaded regions in Fig.~\ref{fig:Chi} show the error bars of
the $\zetap$'s and $\zetaps$'s.
The moments of exits times and the dynamic exponents calculated
from them for all the runs are shown in Appendix~\ref{sec:exit}.
See Appendix~\ref{sec:error} for our method of estimating error. 

The multifractal model~\citep{parisi1985multifractal,frisch} does not
distinguish between doubling and halving times, and hence, for both of them
it predicts the same bridge relation,
$\chi^{\rm{H,D}}_{\rm{p}}= p - \zetap$.
This bridge relation agrees well with our data,
  for the halving times,  (see Fig.~\ref{fig:Chi}),
but not for the doubling times.
  We find unambiguously that $\chiDp\neq\chiHp$, thereby
violating the simple expectation based on the multifractal model.

For the two runs with a solenoidal external force, S1 and S2, we find
$\chiDp = p - \zetaps$, i.e., the bridge relation from the multifractal
model holds for the doubling-time exponents, but with $\zetap$ replaced
by $\zetaps$.
Together this demonstrates that, for the runs S1 and S2,
(a) the solenoidal part of the velocity field is primarily responsible
for the increase of $R$, and (b) both the solenoidal and irrotational parts
affect the decrease of $R$. 
Note that the bridge relations hold for the two values of $\Ma$ we have
used although the precise values of the exponents themselves depend on $\Ma$. 

For the two irrotationally-forced runs, C1 and C2, besides $\chiHp=p-\zetap$,
we find that the relation $\chiHp = p -\zetapc$ also holds because the
scaling exponents, $\zetapc$, of the irrotational
component of the velocity are quite close to $\zetap$. 
Crucially, $\chiDp \neq \chiHp$ again. 
Furthermore, unlike the runs S1 and S2,
$\chiDp \neq p - \zetaps$ (see \Fig{fig:Chi} and \Fig{fig:chi}).
This indicates that the growth of $R$ is not determined solely by the
solenoidal part of the velocity field, but is affected significantly by
the regions of the flows with $\dive\uu > 0$. 
A quantitative understanding of these exponents remains an open question. 
Finally, given the accuracy of our data, we do not detect any dependence of
these exponents on $\Ma$.

In the multifractal model, the velocity field is constructed from
an infinity of interwoven sets each characterised by a scaling exponent $h$
and corresponding fractal dimension $D(h)$. 
It is straightforward to show (see Appendix~\ref{sec:multi} )
that the dynamic exponents $\chip$ and
$D(h)$ are Legendre transforms of each other:
\begin{subequations}
  \begin{align}
&\chip = -\inf_{h}\left[ph+d-p-D(h)\right] \/\\
    &\text{and}\quad D(h) = \inf_{p}\left(ph+d-p+\chip\right)\/.
    \label{eq:leg}
  \end{align}
\end{subequations}
Thus from $\chip$ we determine $D(h)$ -- the dimensionality of structures
that contribute to dynamic scaling\footnote{Even if $D(h)$ is not concave its
  Legendre transform $\chip$ is concave; however, the inverse transform
  returns not $D(h)$ but its concave
  hull~\citep[see, e.g.,][section 8.5]{frisch}.}.
In addition, we must define $\DH(h)$ and $\DD(h)$, which are calculated from
$\chiHp$ and $\chiDp$, respectively, by using Eq.~\eq{eq:leg}. 
This reveals that the halving and doubling times
are influenced by structures of different dimensionality.
All the multifractal spectra are shown in \Fig{fig:Dh}.

We find that external force and $\Ma$ have very little effect on $\DH(h)$
-- in all our runs, $\DH(h) \to 1$  as $h\to 0$. 
Thus the halving times are controlled by shocks, which are
one-dimensional structures characterized by $h\to 0$. 
As expected,  $\DH(h)$ overlaps with the multifractal spectrum
evaluated from equal time exponents $\zetap$.

In contrast, the  function $\DD(h)$ is not universal.
In S1 ($\Ma\simeq 1$), $\DD(h)$ does not go down to $1$, implying that the
structures that influence the scaling of $R(t)$ are more space-filling than
shocks.
In S2 ($\Ma \simeq 2.5$), we find  $\DD(h) \simeq 1$
for $h \simeq 0.4>0$  we infer
that these structures are not shocks but thin elongated patches of high
vorticity, which appear in the vicinity of the shocks,
clearly visible in Fig.\ref{fig:snap}.
Such patches do exist for S1, but they are relatively weaker and broader.
Finally, we note that the function $\DD(h)$ takes values
close to unity in the runs, C1 and C2, as well. 

Note that we have previously undertaken similar studies in simpler
models of compressible turbulence -- the randomly-forced Burgers
equation in one~\citep{De_DynScal} and two dimensions~\citep{De2024Burg2D}. 
They have a few qualitative similarities with the results here -- evidence of
dynamic multiscaling and $\chiDp\neq\chiHp$. 
There are few key differences between the Burgers equation
and the present, more realistic model:
(a) zero pressure (tracers remain trapped on shocks forever),
(b) zero vorticity at all times, and
(c) bifractal velocity fluctuations. 
These additional complexity also precludes the formulation of
building a theoretical framework for the exit times which was possible
in~\citet{De2024Burg2D}.

In summary, in trying to generalize Richardson's law of
pair dispersion to compressible flows, we learn three
important lessons:
(a) Compressible turbulence in two dimensions show dynamic multiscaling;
(b) The dynamic multiscaling obtained here is qualitatively different
from its incompressible counterpart; we need to consider both
doubling and halving times and the two have different dynamic
multiscaling exponents.
The exponents $\chiHp$ satisfies a multifractal model--based bridge
relation and is universal but  $\chiDp$ is not.
They depend on both the external force and the
turbulent Mach number, $\Ma$. 
This non-universality of the inertial--range pair--dispersion statistics is
remarkable, given that the inertial-range fluctuations are supposed to be
independent of the external force.
(c) This dynamic multiscaling \textit{cannot} be understood
within the framework of the present multifractal model of turbulence. 
Consequently, a clear understanding of statistics of pair dispersion in
compressible turbulence remains an open problem, whose solution may
require that we go beyond the multifractal model. 

Furthermore, note that in incompressible turbulence, the Richardson
dispersion can be understood as a scale dependent effective viscosity
and this serves as a measure of mixing by turbulent flows.
We find that in compressible turbulence this description is fundamentally
flawed because we have both doubling and halving times which have
different scaling properties.
As turbulence in most astrophysical context is compressible, mixing in large
scale astrophysical systems, e.g., molecular clouds, must be revisited
in the light of our results.

We speculate that the key qualitative content of our results would be valid in
three dimensions (3D) as well.
In particular, the necessity of the two kinds of exit
times, dynamic multiscaling, and intermittency of pair dispersion
should carry over to 3D. 

Highly supersonic turbulent flows govern various astrophysical processes,
e.g. star formation in molecular clouds, and the transport and mixing of gases
along with their chemical kinetics in the interstellar
media~\citep{Mordecai2004Review,Elmegreen2004Review_ISM}. 
Our work sets out an effective paradigm which can potentially provide in-depth
insights into these aspects.

\begin{acknowledgments}
  DM and RP thank the Isaac Newton
  Institute for Mathematical Sciences, Cambridge, for support and hospitality
  during the programme Anti-diffusive dynamics: from sub-cellular to
  astrophysical scales, where partial work on this paper was undertaken.
  This work was supported by EPSRC grant EP/R014604/1.

  DM acknowledges the support of the Swedish Research
  Council Grant No. 638-2013-9243. Nordita is partially supported by NordForsk.
  S.D. thanks the PMRF (India) for support. R.P. and S.D. thank the Science and
  Engineering Research Board (SERB), 
  Anusandhan National Research Foundation (ANRF) and the 
  National Supercomputing Mission (NSM),
  India for support, and the Supercomputer Education and Research Centre (IISc)
  for computational resources.
\end{acknowledgments}
\bibliography{compturb}
\clearpage
\appendix
\onecolumngrid
\section{Numerical simulation}
\label{sec:numsim}
The governing equations, in vectorial notation, are as follows:
\begin{widetext}
\begin{subequations}
\begin{align}
    & \frac{\partial\rho}{\partial t} + \grad\cdot(\rho\uu)=0\,,
    \label{eq:Cont1app}
    \\
    & \frac{\partial\uu}{\partial t} + \uu\cdot\grad\uu =  [2\nu{\bm S}+{\bm I}_2(\zbulk\dive\uu-\cs^2)]\cdot\grad(\ln\rho) +
    \nu\grad^2\uu + \dive(\zbulk\dive\uu) +  \ff_{\rm ext}\,.
    \label{eq:Mom1app}
\end{align}
\end{subequations}
\end{widetext}
Here $\uu(\xx,t)$ is the velocity field,
$\rho(\xx,t)$ is the density field,
$P$ is the pressure,
$\nu$ is the kinematic shear viscosity
taken to be a constant~\citep{Haugen2022ClusteringInertial,
  Federrath2011Machno, Thibaud2022Dissipation}.
$\zbulk$ is the bulk viscosity,
$\delta_{ij}$ is the Kronecker delta,
$S_{ij} \equiv (1/2)\left(\delj \ui + \deli \uj\right) -
           (1/2)\delta_{ij}\delk\uk$ 
is the traceless rate-of-strain tensor,
and the equation of state is given by 
$P = \cs^2\rho$, where $\cs$ is the (constant) speed of sound. 
Here, $f_{{\rm ext}\,,i}=\mathcal{P}_{ij}f_j$ and
${\bm I}_2$ is the $2\times2$ identity matrix.
In 2D, Eq.~\eqref{eq:Mom1app} can be rewritten in terms of two scalars -- the
vorticity, $\omega=\zhat\cdot\curl\uu$, and the divergence of velocity,
$\mathcal{D}=\dive\uu$.

We use a pseudo-spectral code to solve \eq{eq:Mom1app} rewritten
in $\omega$ and $\mD$. 

\subsection{De-aliasing}
\label{sec:dealias}
To remove aliasing errors, we use an exponential
filter~\citep{Hou2007ExponentialFilter}:
\begin{equation}
    \mathcal{F}(k) = \exp\left[-36.0\left(\frac{k}{\km}\right)^{36.0}\right]\,,
    \label{eq:filter}
\end{equation}
where $\km=N/2$ is the highest allowed wave-number in the simulation,
$N$ the number of grid points in a given dimension. 
The filter $\mathcal{F}(k)$ is approximately unity up to $k\approx N/3$ and
drops sharply but smoothly to almost zero beyond that, essentially all modes
with $k>N/3$ are strongly damped.
It is a better choice than the Galerkin-based discrete cut-off filter
which are usually used for incompressible turbulence because it reduces
oscillations around shocks which result from
Galerkin-trucation, enabling us to simulate supersonic turbulence at
high Mach numbers while maintaining relatively large Reynolds numbers.

\subsection{External force}
\label{sec:force}
The quantity, $\fvec(\xvec,t)$, is a white-in-time random noise. 
Its spectrum in the spatial domain is restricted to large spatial scales
(small $k$). 
In our simulations, we use
\begin{subequations}
  \begin{align}
    \widehat{\fvec}_{k_x}(\kvec,t_n) &= A  \cos(\theta_{\kvec,n})e^{i\phi_{\kvec,n}}\/,\  \\
\text{and}\quad    \widehat{\fvec}_{k_y}(\kvec,t_n) &= A \sin(\theta_{\kvec,n})e^{i\phi_{\kvec,n}}\/,
  \end{align}
\end{subequations}
\label{eq:cNSForceSim}
where the amplitude,
\begin{equation}
    A = \begin{cases}
    f_0\sqrt{\delta t} &\text{for}\quad k=\kf \,; \\
    0 &\text{otherwise}\,.
    \end{cases}
    \label{eq:Forc_amp}
\end{equation}
$f_0$ is constant, and $\theta_{\kvec,n}$ and $\phi_{\kvec,n}$ are uniformly
distributed random numbers belonging to the interval $[0,2\pi]$. 
We choose $\kf=3$. 
The ratio of power being injected into the solenoidal and irrotational modes
is tuned by applying the following projection operator,
\begin{equation}
    \hat{\mathcal{P}}_{ij}=\xi\delta_{ij}+ (1-2\xi)k_ik_j/k^2\,,
    \label{eq:Pij_cNS}
\end{equation}
which yields $\fvec_{\rm ext}(\xx,t)$.
\subsection{The shock viscosity $\zbulk$}
\label{sec:shock}
The shock viscosity, $\zbulk$, is a spatially non-uniform bulk viscosity which
is locally enhanced in the neighbourhood of shocks and serves to resolve the
strong shocks on a
finite-resolution grid in the high Mach number simulations. 
We evaluate $\zbulk$ on the basis of the prescription outlined in
Ref.~\cite{vonneumann1950method}. 
At each grid $(i,j)\equiv\xvec$,
\begin{equation}
    \zbulk(\xvec) = c_{\rm shock}\avg{\overline{\mathcal{D}}}_5\delta x \delta y\,,
    \label{eq:shockvis}
\end{equation}
where $\delta x = \delta y=L/N$ are the grid-spacings along each direction, and $\avg{...}_5$ denotes the average over a $5\times5$ domain centred at $(i,j)$.
\begin{equation}
    \overline{\mathcal{D}} = \begin{cases}
    |\dive\uu| &\text{for}\quad \dive\uu<0 \,; \\
    0 &\text{otherwise}\,.
    \end{cases}
    \label{eq:Dbar}
\end{equation}
We choose $c_{\rm shock}\sim\mathcal{O}(1)$. 
Thus, $\zbulk$ is non-zero around shocks, decreasing quickly with distance from them, and zero elsewhere. 

\subsection{Timestepping}
\label{sec:time}
We implement the white-in-time external force with Euler Marayuma
scheme~\citep{higham2001algorithmic} 
and exponential Adams-Bashforth method~\citep{cox2002exponential} for
timestepping the rest of the equation. 
\subsection{Lagrangian particle tracking}
The position of the $\rm{m}$-th tracer, $\XXm(t)$,
evolves according to
\begin{subequations}
  \begin{align}
\frac{d\XXm(t)}{dt}&=\Vvec^{\rm m}(t) \\
\text{where}\quad \Vvec^{\rm m}(t)&=\uu(\xx,t)\delta(\xx-\XXm)\/.
\end{align}
\end{subequations}
The velocity at the off-grid position of a tracer is obtained by linear
interpolation from the neighboring grid points and the timestepping
is done by a Runge-Kutta scheme. 
\section{Results}
\subsection{Parameters and Snapshots}
\label{sec:para}
\begin{figure*}
    \centering
    \includegraphics[width=0.99\textwidth]{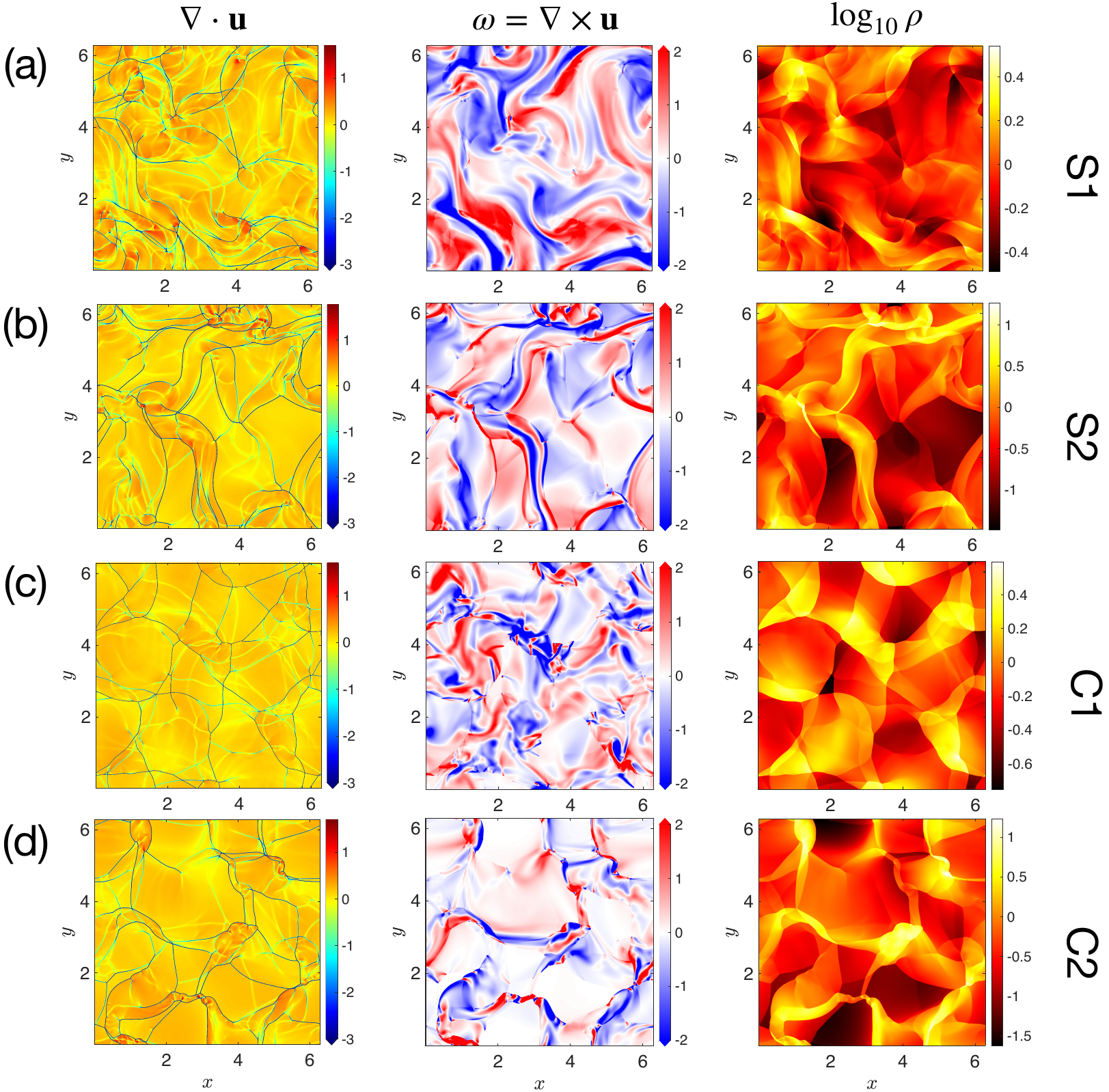}
    \caption{\small \textbf{Typical snapshots} of  $\divu$,
      vorticity $\omega=\zhat\cdot\curl\uu$, and
      $\log_{10}\rho$, in the cases (a) S1, (b) S2, (c) C1 and (d) C2.
      In each plot, $\dive\uu$ and $\omega$ are normalized by their respective
      root-mean-square (rms) values. 
      Shocks are clearly visible in the $\dive\uu$ profiles as the blue
      filament--like structures. 
      The vorticity field, $\omega$, is smoother in S1 than in S2 which contains
      nearly one-dimensional structures of intense vorticity bordering the
      shocks; 
      in all figures, the areas of high vorticity lie in the vicinity of shocks 
      Similarly the spatial distribution of $\omega$ is more
      homogeneous in C1 than in C2. 
      The values of $\rho$ are clearly enhanced near the shocks; for a given
      type of external forcing, $\rho$ tends to take much higher values at
      higher Mach numbers, $\Ma$.      
    }
    \label{fig:Snap}
\end{figure*}

The key parameters of our DNS are summarized in Table~\ref{tab:para}.
In Fig.~\ref{fig:Snap} we show representative snapshots  of
 $\dive\uu$, $\oo = \curl\uu$, and $\log_{10}\rho$ for all runs. 
The shocks are clearly visible as filamentary structures with large
negative values of $\dive\uu$.
Naively, we may expect that the presence of the shocks would not
show up in pseudocolour plot of $\oo$, but this not the
case -- the shocks can be seen as almost one-dimensional
manifolds with large absolute values of the vorticity: 
This is because, although a shock front that is a straight line does
not contribute to vorticity, \textit{curved} shock fronts do. 
In our simulations,
(a) the density fluctuations are larger for the supersonic cases
compared to the transonic ones and
(b) the vorticity is more uniformly distributed in the transonic
cases than the supersonic ones, for a given type of external forcing,
as observed in Fig.~\ref{fig:Snap}.
\begin{table*}
  \begin{center}
\def~{\hphantom{0}}
  \begin{tabular}{ccccccccccccc}
      \hline
      Run & $\xi$ & $\cs$ & $\urms$ & $\Ma$ & $\Rey$ & $\LI$ & $\TE$ & $\Psi$ & $\epsilon$ & $\lambda$ & $\Rey_\lambda$ & $\eta$
      \\[3pt]
      \hline
      \hline
       S1 & 1 & 0.05 & 0.053 & $1.05$ & $6.7\times10^3$ & 1.66 & 32 & $0.82$ & $8.1\times10^{-6}$ & 0.29 & 306 & 0.01 \\
       S2 & 1 & 0.02 & 0.050 & $2.5$ & $6.3\times10^3$ & 1.20 & 24 & $0.68$ & $9.6\times10^{-6}$ & 0.26 & 255 & 0.01 \\
       C1 & 0 & 0.05 & 0.038 & $0.75$ & $4.7\times10^3$ & 1.04 & 28 & $0.19$ & $1.3\times10^{-5}$ & 0.16 & 123 & 0.01 \\
       C2 & 0 & 0.02 & 0.040 & $2.0$ & $5.0\times10^3$ & 1.17 & 29 & $0.34$ & $8.7\times10^{-6}$ & 0.21 & 172 & 0.01 \\
       \hline
  \end{tabular}
  \caption{\small \textbf{Simulation parameters.} Runs S1 and S2 have a 
    solenoidal external force whereas C1 and C2 have an irrotational force.
    The parameters are defined as follows:
    $\cs$ is the speed of sound,
    $\urms$ is the root-mean-square velocity,
    $\Ma$ is the Mach number,
    $\Rey=(\urms L/\nu)$ is the box-size Reynolds number,
    $E(k)$ is the kinetic energy spectrum,
    $\LI=\pi\sum_{k}[E(k)/k]\Big/\sum_{k}E(k)$ is the integral length scale,
    $\TE=\LI/\urms$ is the large-eddy turnover time,
    $\Psi$ is the solenoidal ratio defined in \eq{eq:Psi},
    $\epsilon=\avg{\nu\rho \sigma_{ij}\partial_ju_i}\Big/\avg{\rho}$
    is the rate of dissipation of kinetic energy per unit mass,
    $\lambda=\sqrt{5\nu\urms^2\Big/\epsilon}$ is the Taylor microscale,
    $\Rey_\lambda=\urms\lambda\big/\nu$ is the Taylor-microscale Reynolds number,
    and $\eta=\left(\nu^3\big/\epsilon\right)^{1/4}$ is the dissipation scale. 
    In all the runs, $\nu=5\times10^{-5}$ and number of tracers is
    $n=4096^2$. Note that the length scales in this Table are not
    normalized by $L=2\pi$.}
  \label{tab:para}
  \end{center}
\end{table*}

\subsection{cascade and spectra} 
\label{sec:spec}
\begin{figure*}
    \centering
    \includegraphics[width=0.9\textwidth]{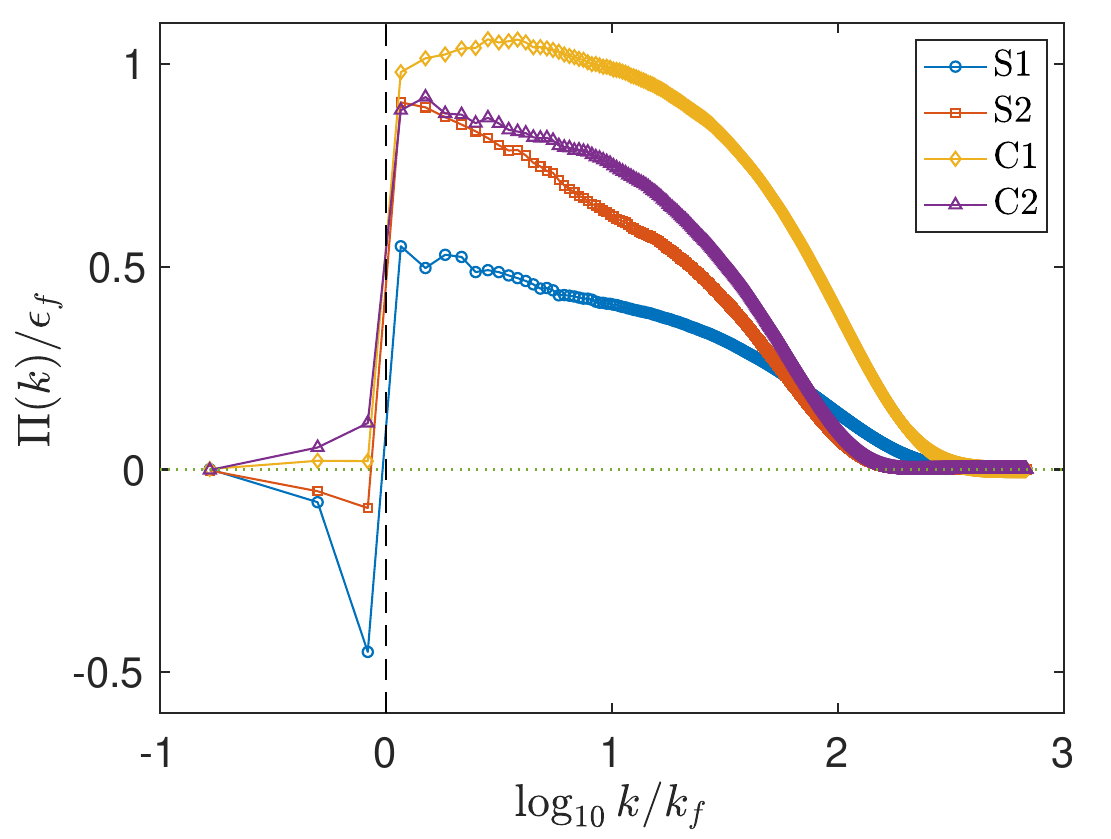}
    \caption{\textbf{Kinetic energy flux} for the four DNS runs.
    We obtain a range of direct cascade.  }
    \label{fig:flux}
\end{figure*}
\begin{figure*}
    \centering
    \includegraphics[width=0.9\textwidth]{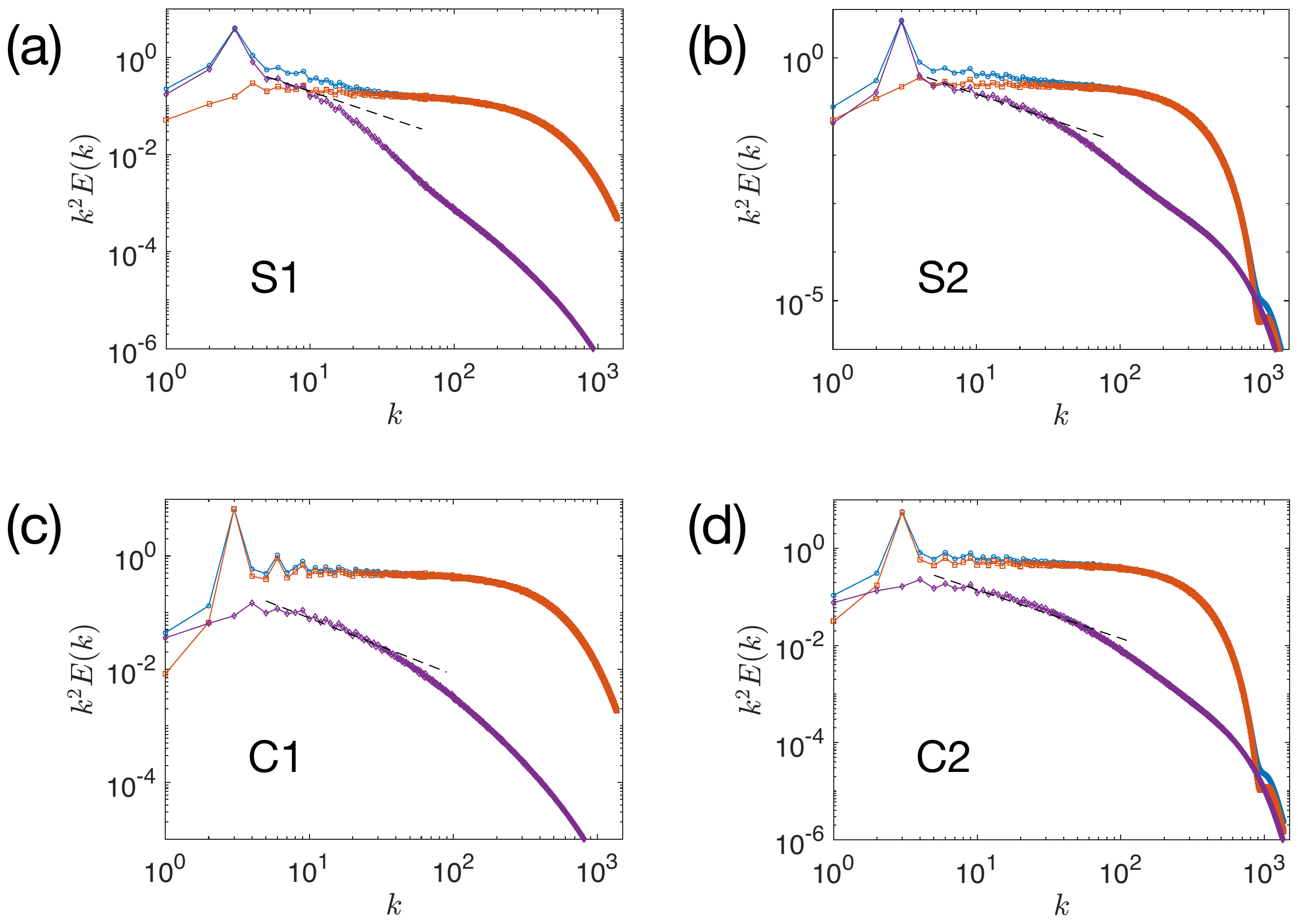}
    \caption{\textbf{Compensated velocity spectra} Log-log plots of the
      compensated (by $k^2$) spectra of the total velocity (blue),
      $E(k)$, its compressive component
      (red), $\Ec(k)$, and its solenoidal components (violet), $\Es(k)$
      for the different runs.
      The spectra are normalized by $\sum_k E(k)$. }
    \label{fig:Espec}
\end{figure*}
Using the Helmholtz decomposition, we write
$\uu =\us+\uc$, where $\us$ and $\uc$ are, respectively,
the solenoidal ($\nabla\cdot\us=0$) and irrotational ($\nabla\times\uc=0$)
components of the flow.
The relative importance of $\us$ and $\uc$ in the total
kinetic energy is measured by the \textit{solenoidal ratio},
\begin{equation}
  \Psi\equiv \frac{\avg{\rho|\us|^2}}{\avg{\rho|\uvec|^2}}\/,
\label{eq:Psi}
\end{equation}
which is the ratio of the total kinetic energy in the solenoidal modes to the
total kinetic energy.

In two-dimensional \textit{incompressible} turbulence, forcing at an
  intermediate scale gives rise to two cascades: (a) an inverse cascade
  to large scales and (b) a direct cascade to small scales.
  The statistics in the inverse cascade is close to Gaussian, whereas
  the statistics in the direct cascade is intermittent. 
  In two dimensional \textit{compressible} turbulence, forced at intermediate
  scales, kinetic energy shows a dual cascade~\citep{falkovich2017vortices}.
  We choose to force at the large scales, $\kf = 3$, such that
  we obtain a large range of direct cascade, see \Fig{fig:flux}.
  The statistics of velocity fluctuations in this range is
  intermittent, as we show later by evaluating the scaling exponents of
  the $p$-th order structure functions. 

We display the three shell-averaged velocity spectra,
\begin{subequations}
  \begin{align}
    E(k)&=|\hat{\uvec}(\kvec)|^2\/,\\
    \Es(k)&=|\hat{\us}(\kvec)|^2\/, \\
   \text{and}\quad \Ec(k)&=|\hat{\uc}(\kvec)|^2
  \end{align}
\end{subequations}
respective for the total, solenoidal, and compressive velocity.
They are plotted in \Fig{fig:Espec}. 
In all the runs, $\Ec\sim k^{-2}$ over a significant range of scales.
Earlier works~\citep{falkovich2017vortices} have also observed this
over a smaller range of scales and have identified it with
acoustic turbulence.
In S1, $\Es(k)$ falls off much faster than that in S2. 
Thus, the small-scale fluctuations of $\us$ are much smoother in S1 than
those in S2, and this is a reflection of the elongated intense vorticity
patches, which are aplenty in S2, but not in S1. 
In S2, $\Es(k)$ appears to scale as $k^{-3}$, across a small regime of $k>\kf$. 
In the irrotationally-forced runs, $\Ec(k)\gg E_s(k)$ across whole of the
inertial range with the difference being small at higher $\Ma$.  
$\Es(k)$, however, does not appear to have a well-defined power-law regime.
\subsection{Equal-time structure functions}
\label{sec:stfun}
\begin{figure*}
    \centering
    \includegraphics[width=0.9\textwidth]{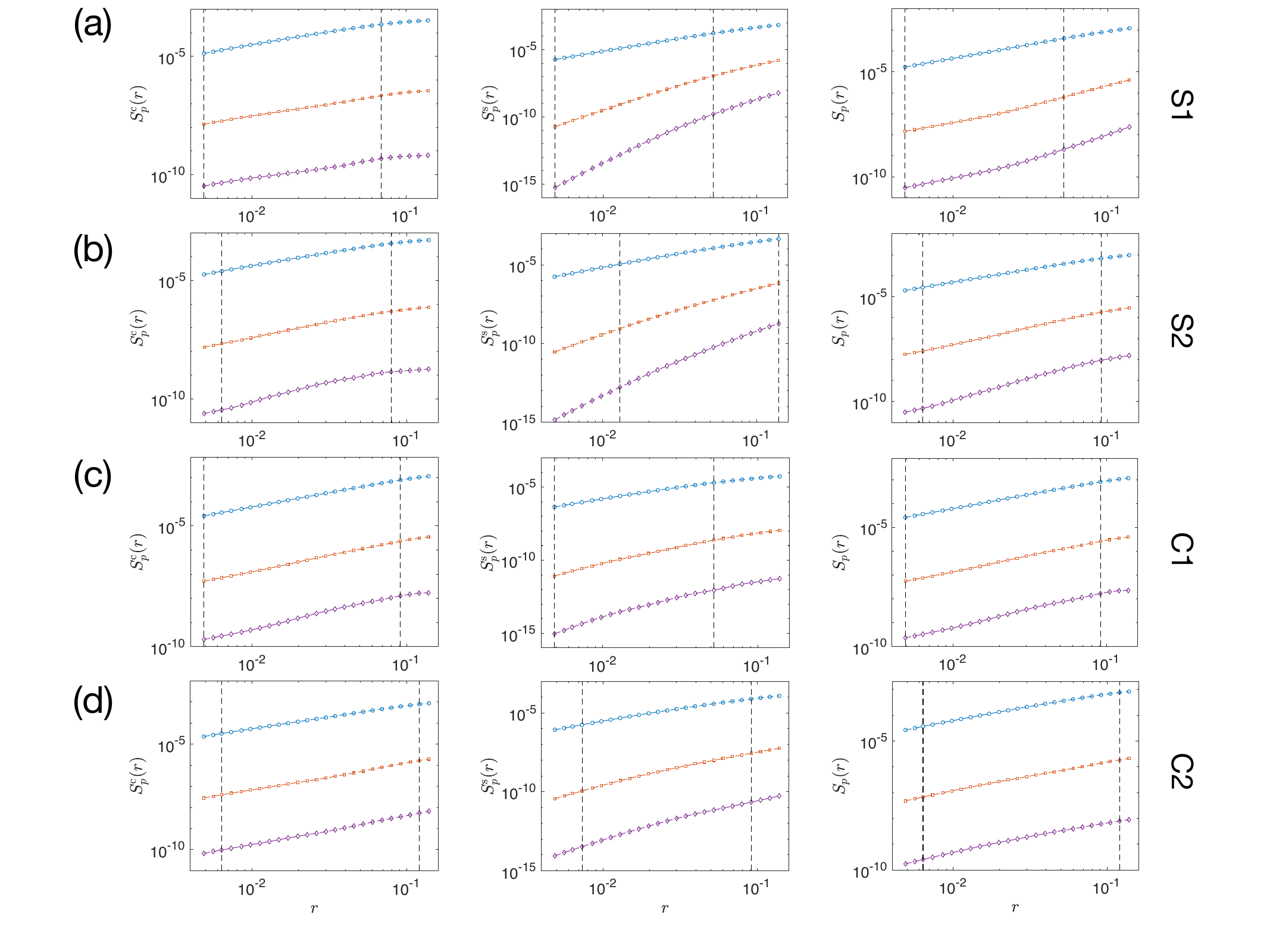}
    \caption{\small \textbf{Equal-time structure functions.} Log-log plots of $\Sp(r)$, $\Sps(r)$ and $\Spc(r)$, defined in \eq{eq:spr} versus $r$ for
      $p=2$ (blue), $p=4$ (red) and $p=6$ (violet) in runs
      (a) S1, (b) S2, (c) C1 and (d) C2. 
    The vertical dashed lines demarcate the region of power-law fitting in each case across which we extract $\zetap$, $\zetaps$ and $\zetapc$.
    In each figure, the data has been averaged over a single non-equilibrium steady state (NESS) snapshot. 
    }
    \label{fig:eqstr}
\end{figure*}
\begin{figure*}
    \centering
    \includegraphics[width=0.9\textwidth]{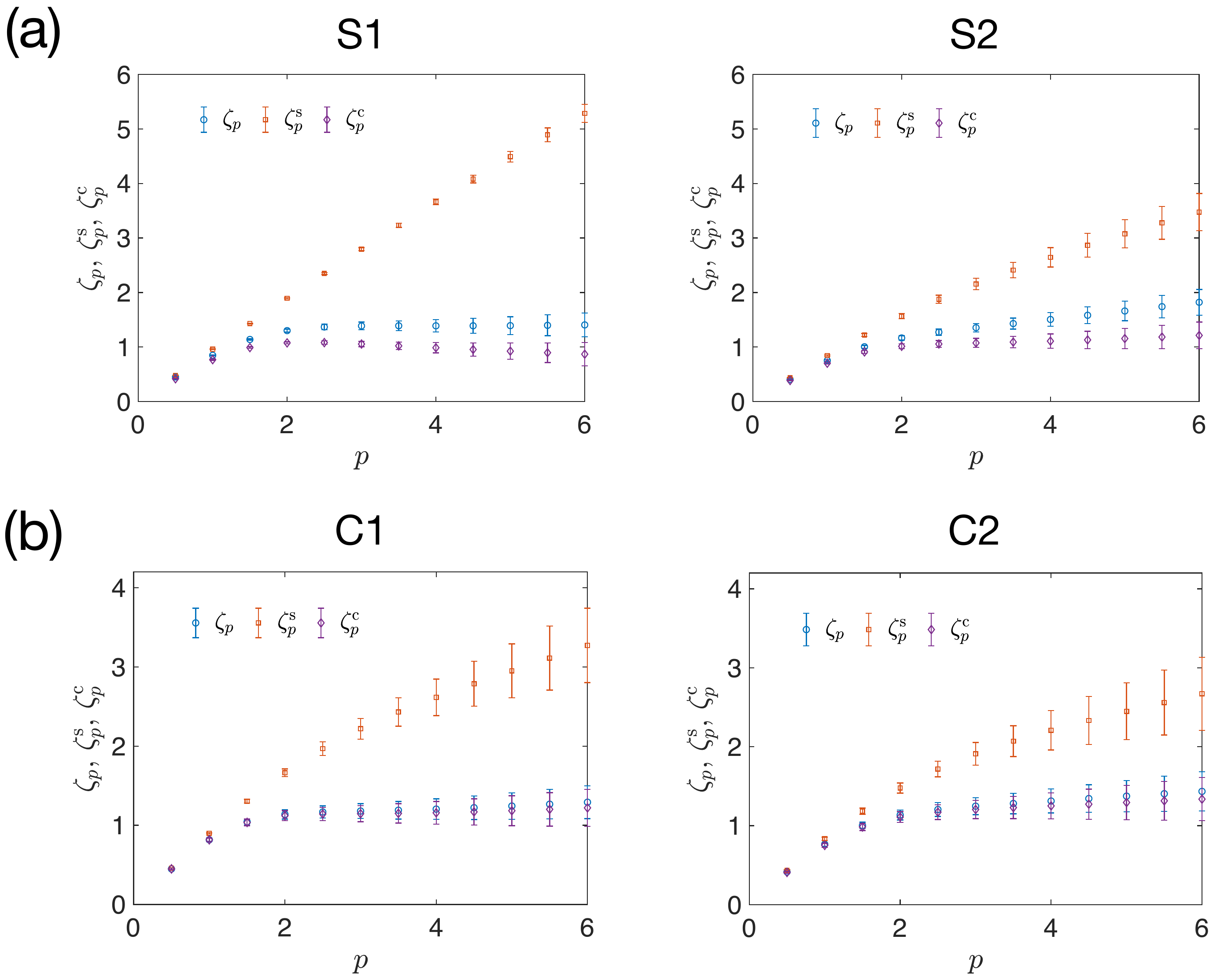}
    \caption{\small \textbf{Equal-time exponents $\zetap$}
      for the four runs. }
    \label{fig:zetap}
\end{figure*}
In compressible turbulence, we can define $p$-th order longitudinal structures
functions of $\uu$ as well as its components, $\us$ and $\uc$:
\begin{widetext}
\begin{subequations}
  \begin{align}
    \delta u_{\parallel}(r) \equiv
               \left[\uu(\xx+\rr)-\uu(\xx)\right]\cdot\frac{\rr}{r}\/\quad &
    \text{and}\quad\delta u_{\parallel}^{\rm{s,c}}(r) \equiv
               \left[\uu^{\rm{s,c}}(\xx+\rr)-\uu^{\rm{s,c}}(\xx)\right]
               \cdot\frac{\rr}{r}\/; \\
     \text{with}\quad\Sp(r)=\avg{| \delta u(r)_{\parallel} |^p}\/,
               \quad &
      \Sp^{\rm s,c}(r)=\avg{|\delta u(r)^{\rm{s,c}}_{\parallel} |^p}\/,
                  \\
      \text{and}\quad \Sp(r)\sim r^{\zetap}\/,\quad & \Sps(r)\sim r^{\zetaps}\/,
                          \quad \Spc(r)\sim r^{\zetapc}\/.
    \end{align}
    \label{eq:spr}
\end{subequations}
\end{widetext}
The scaling relations hold for $\eta\ll r\ll \LI$, where $\eta$ is the
dissipation scale and $\LI$ is the integral scale. 
All the exponents, $\zetap$, $\zetaps$ and $\zetapc$ are
nonlinear monotone increasing functions of
$p$~\citep{Konstandin2012EulLag,wang2012scaling,wang2017scaling}. 

We show plots of various equal-time longitudinal structure functions of the
total velocity, $\Sp(r)$, and its solenoidal, $\Sps(r)$, and irrotational,
$\Spc(r)$, components in \Fig{fig:eqstr}. 
In each image, the regime of power-law fitting has been demarcated by
dashed lines. 
In Figs.~\ref{fig:eqstr}(a) and \ref{fig:eqstr}(b), note that the scaling
regime of $\Sps(r)$ in S2 occurs across higher values of $r$ than that in S1. 
Moreover, for a given $p$, the slope in the scaling-range of $\Sps(r)$ in the
former is visibly smaller than that of $\Sps(r)$ in the latter. 
For $S_{\rm 2}^{\rm s}(r)$, the nature of this change can also be
inferred from the scaling of $\Es(k)$ in \subfig{fig:Espec}{a} and
\subfig{fig:Espec}(b). 
This is indicative of the appearance of the strip-like intense vorticity
patches in S2. 
\subsection{Pair dispersion}
\label{sec:pair}
\begin{figure*}
    \centering
    \includegraphics[width=\textwidth]{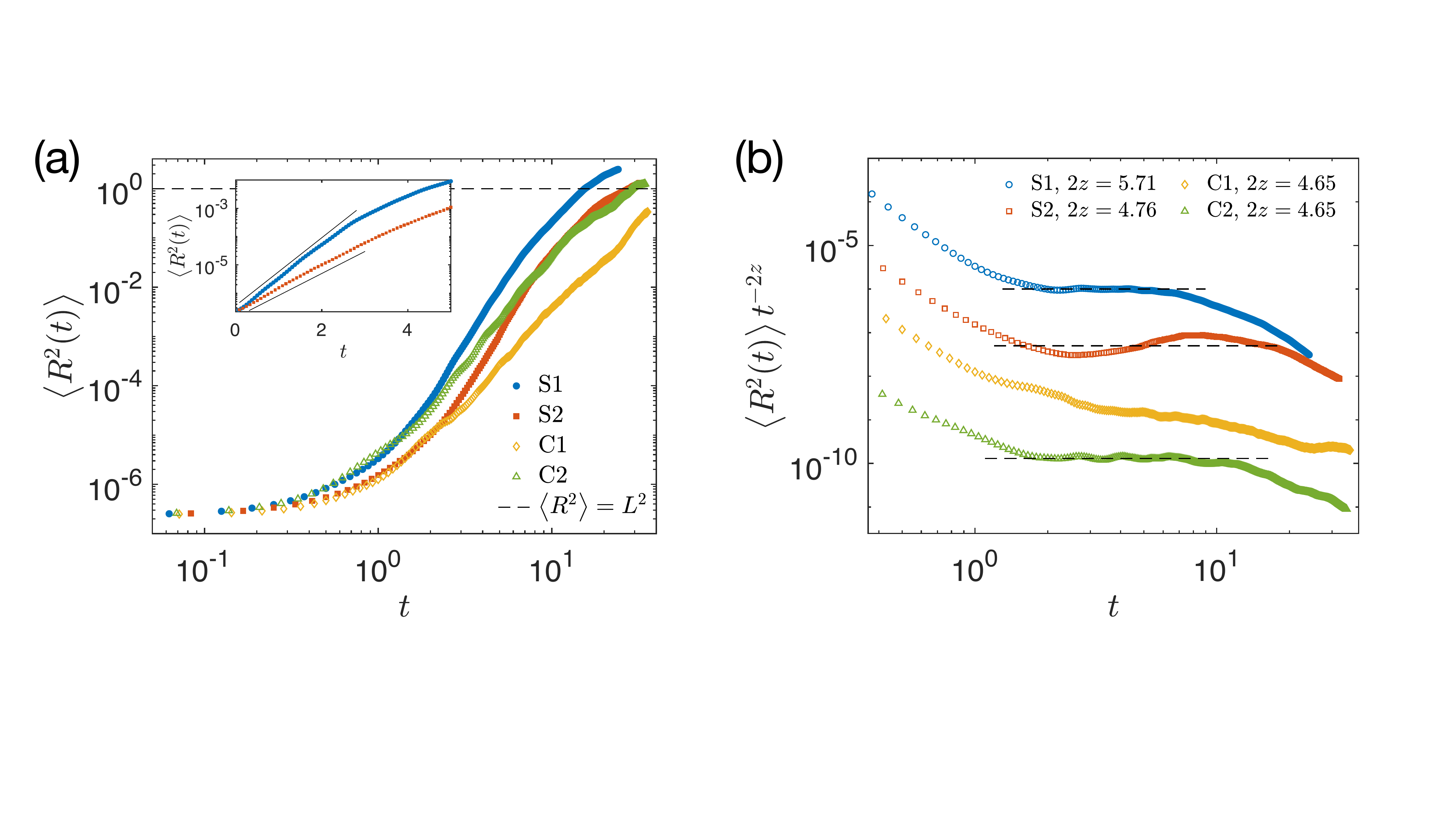}
    \caption{\small \textbf{Mean-squared separation.} (a) Log-log plots of the
      time evolution of the mean-squared separation, $\avg{R^2(t)}$, between
      two tracers.  
      The growth of $\avg{R^2}$ is faster in S1 than in S2, and faster in C2
      than in C1. 
      \textit{Inset:} Semi-log graphs of $\avg{R^2}$ for S1 and S2 displaying
      exponential growth at early times.
      (b) Log-log plots of $\avg{R^2}$ compensated by $t^{2/z}$,
      where $z=1-\zetat/2$; 
      power-law growth of $\avg{R^2}$ across its inertial range values appears
      to occur in every run; in C1, $\avg{R^2}$ clearly scales more slowly
      than $t^{2/z}$ in this range.
    }
    \label{fig:Rsquare}
\end{figure*}
  We calculate how the separation,
  $R(t)$, between a pair of nearest-neighbor tracers, separated initially by
  the grid-spacing (approximately $0.3\eta$), evolves with time.
  We obtain $\avg{R^2(t)}$ by averaging, at every instant, over the
  $2n$ separations. 
  Our results are shown in \subfig{fig:Rsquare}{a}.
  At early times, $\avg{R^2(t)}$ grows exponentially for all the
  runs [For S1 and S2 we show the plots in the inset of
    \subfig{fig:Rsquare}{a}].
  This range of exponential growth is very small for C1 and C2. 
Similar results~\citep{Jullien2003DipsersionExp} have been obtained for
incompressible 2D turbulence.
Exponential growth up to a length scale typically implies that the flow across
the tracers is smooth up to that scale. 
This holds for our simulations because at such small time scales most of the
pair separations lie away from shocks.
We note that at late times the behavior of $\avg{R^2}$ is
not universal.

If $R(t)$ lies in the inertial range, then
$R^2 \sim t^{2/(1-h)}$, where $h$ is the local H\"older exponent of the velocity
fluctuations in space. 
If we use Kolmogorov's simple scaling hypothesis, $\zetat=2h$ which yields
$\avg{R^2(t)} \sim t^{2/z}$, where $z=1-\zetat/2$.
In incompressible flows, $\zetat = 2/3$ and $z=2/3$, which
corresponds to Richardson scaling.
For our simulations, in each case, we calculate $\zetat$
from the scaling of the second order structure
functions, $S_{\rm 2}(r)$.
In \subfig{fig:Rsquare}{b}, we display the plots of $\avg{R^2}$ compensated by
$t^{2/z}$. 
It appears that there is a range of scales over which $\avg{R^2}\sim t^{2/z}$
in S1, S2 and C2, while the increase in $\avg{R^2(t)}$ in C1 is visibly
  slower that $t^{2/z}$.
However, in every case, the scaling regime is too small to make robust
  quantitative measurements of the values of $z$ from the $\avg{R^2(t)}$
  versus $t$ plots. 

\subsection{Exit times}
\label{sec:exit}
\begin{figure*}
    \centering
    \includegraphics[width=0.8\textwidth]{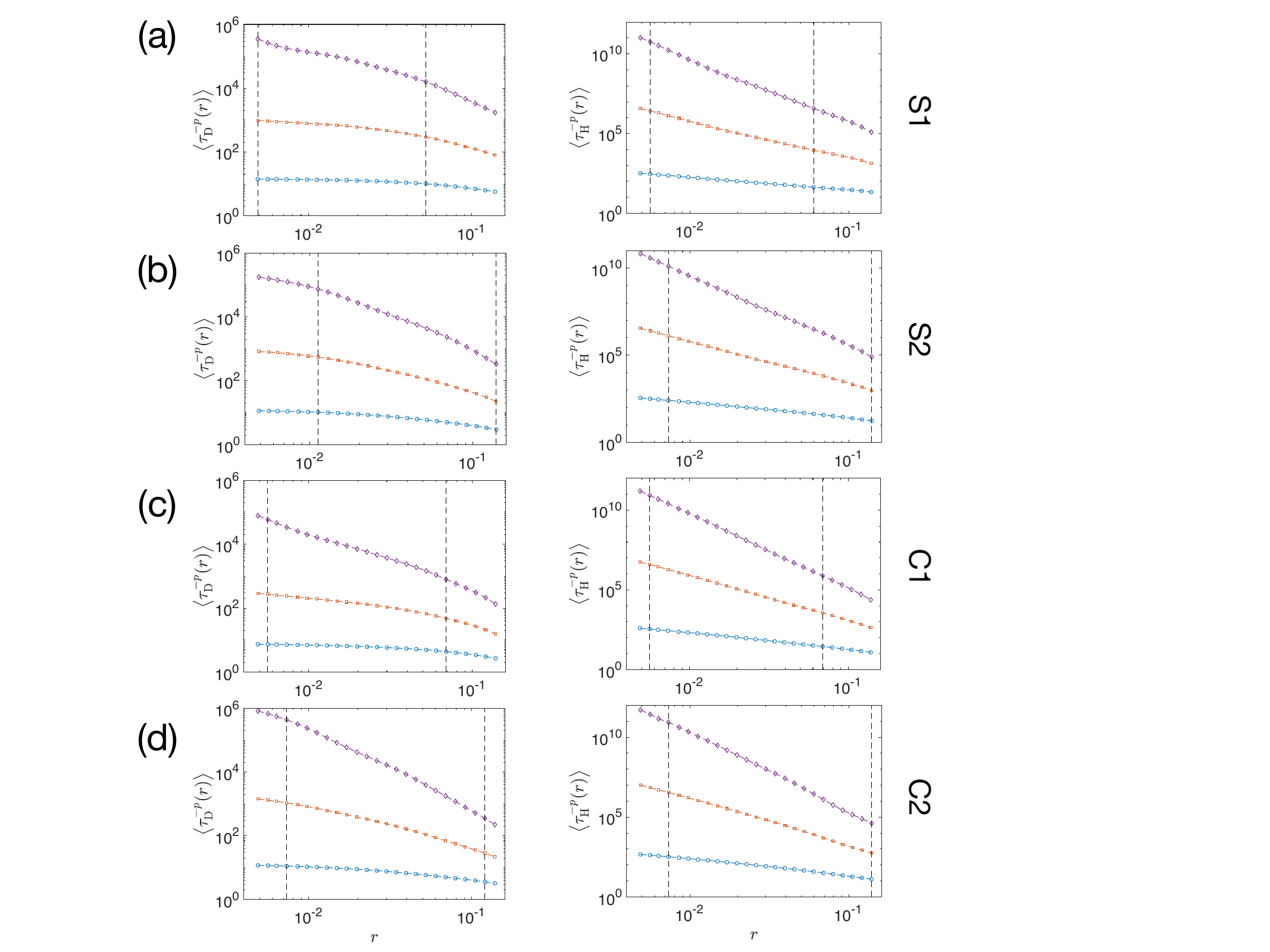}
    \caption{\small \textbf{Moments of Lagrangian exit times.} Log-log plots of
      $\avg{\tauD^{-p}(r)}$ and $\avg{\tauH^{-p}(r)}$ versus $r$ for
      $p=2$ (blue), $p=4$ (red) and $p=6$ (violet) in runs
      (a) S1, (b) S2, (c) C1 and (d) C2. 
      The vertical dashed lines demarcate the region of power-law fitting in
      each case across which we extract $\chiHp$ and $\chiDp$.}
    \label{fig:Texits}
\end{figure*}

The graphs showing the scaling of the moments of the Lagrangian exit times,
$\avg{\tauD^{-p}(r)}$ and $\avg{\tauH^{-p}(r)}$, are given in
Fig.~\ref{fig:Texits}.
\begin{figure*}
    \centering
    \includegraphics[width=0.8\textwidth]{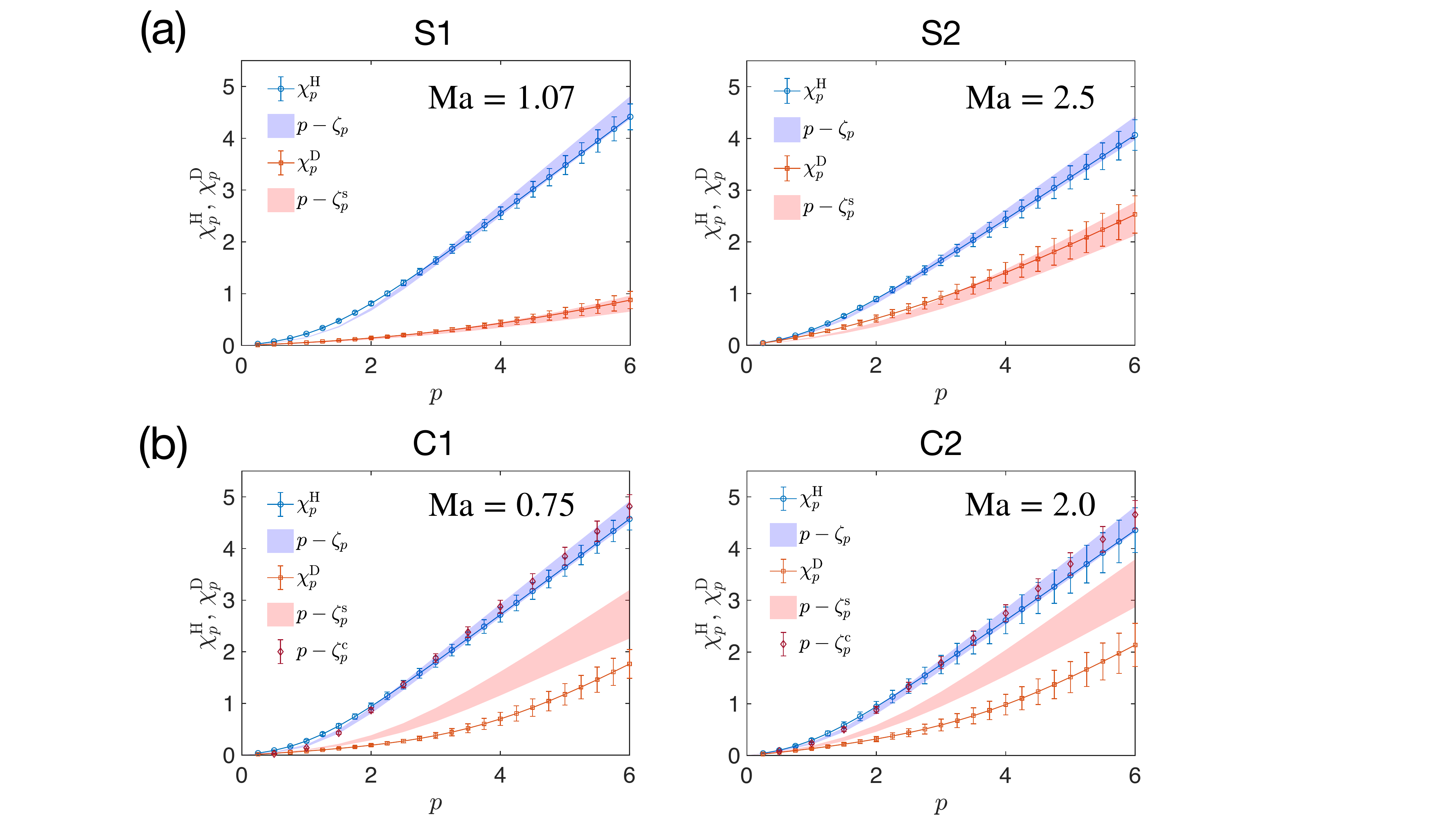}
    \caption{\small \textbf{Dynamic exponents} for doubling and halving times
      $\chiDp$ and $\chiHp$, respectively, as functions of the
      order $p$ which takes integer and fractional values up to $6$. 
      The blue shaded region corresponds to the bridge relation, $p-\zetap$,
      derived from the multifractal model.
      The pink shaded region corresponds to the expression $p-\zetaps$. }
    \label{fig:chi}
\end{figure*}
\subsection{The multifractal spectrum}
\label{sec:multi}
\begin{figure*}
    \centering
    \includegraphics[width=\textwidth]{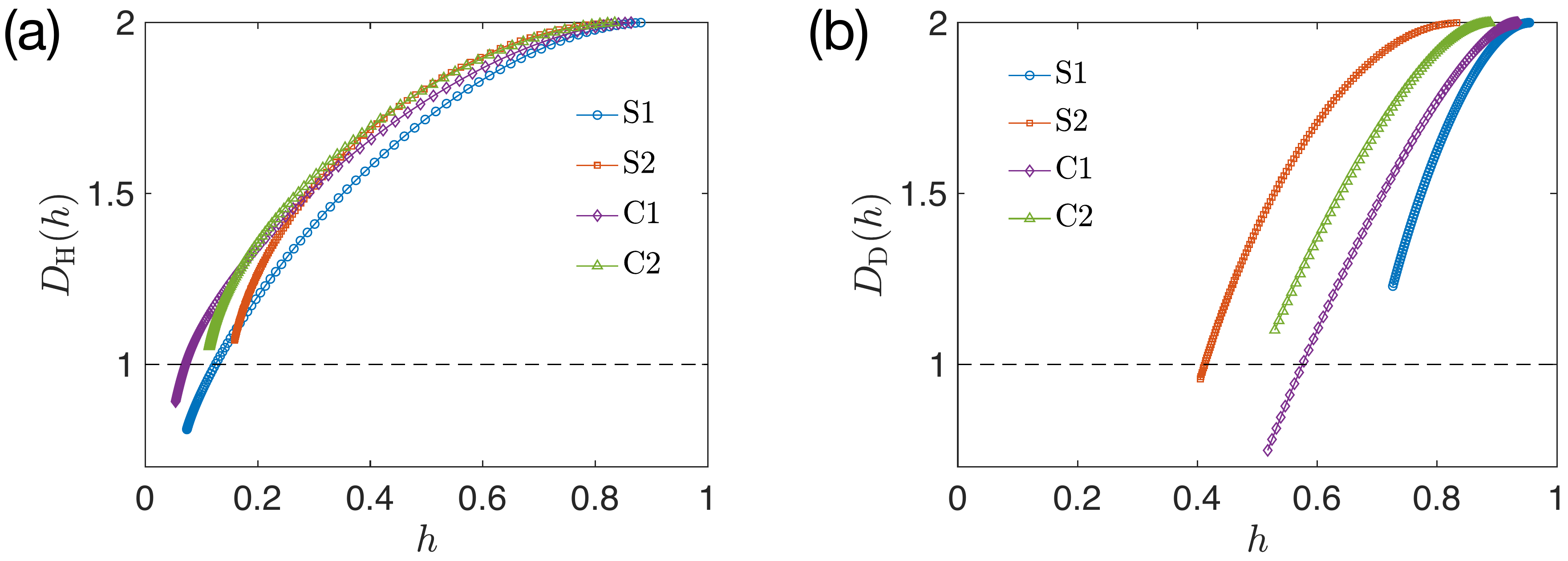}
    \caption{\small \textbf{Multifractal spectrum}. Plots of (a) $\DH(h)$  and
    (b) $\DD(h)$ versus $h$ for all the runs. 
   $\DH(h)$ changes slightly as we change the type of external force and $\Ma$; 
     the dependence of $\DD(h)$ on $\Ma$ is much stronger in case of the solenoidal external force than the irrotational one. 
     For S2, C1 and C2, $\DD(h)$ takes values close to unity but not S1, implying that 1D structures contribute to the scaling of the doubling times in the former cases.
     }
    \label{fig:Dh}
\end{figure*}
Following the  multifractal model we consider the velocity field to comprise
an infinity of interwoven sets, $\Sh$, each characterized by a scaling
exponent, $h$, and fractal dimension, $D(h)$. 
For any point $\xx\in\Sh$, $\dup(R) \sim R^h$, in the limit $R/\LI\to 0$.
We use this model to calculate the dynamic exponents, $\chip$. 
The characteristic time scale (here, exit-time),
$\tauxx(R) \sim R/[\dup(R)]\sim R^{1-h}$.
Its moments of negative orders, $-p$ ($p>0$), scale as
\begin{subequations}
  \begin{align}
    &\avg{\tau^{-p}(R)} \sim \int dw(h)
    \left(\frac{R}{\LI}\right)^{d-D(h)-p(1-h)}
  \sim R^{-\chip}\/; \label{eq:mf}\\
  &\implies -\chip = \inf_{h}\left[ph+d-p-D(h)\right] \/\\
  &\text{and}\quad D(h) = \inf_{p}\left(ph+d-p+\chip\right)\/.
  \label{eq:Leg}
  \end{align}
  \label{eq:dynscal}
\end{subequations}
The spatial dimensionality $d=2$, and $dw(h)$ gives the weights of the different
exponents.
For $R/\LI\to 0$, the integral in \eq{eq:mf} is evaluated by the method of
steepest descents. 
Equation~(\ref{eq:Leg}) implies that $D(h)$ and $\chip$
are Legendre transforms of each other. 
Physically, $D(h)$ reveals the dimensionality of structures that contribute to
the scaling of $\dup(R)$ with the exponent $h$, unveiling the
\textit{multifractal spectrum}. 
Note that, even if $D(h)$ is not concave its Legendre transform $\chip$
will be concave; however, the inverse transform returns not $D(h)$ but its
concave hull~\citep[see, e.g.,][section 8.5]{frisch}. 
\subsection{Error analysis}
\label{sec:error}
For equal-time structure functions we take $M=10$, well-separated-in-time,
snapshots, from the stationary state of our DNS.
From each snapshot we calculate the structure function~\eq{eq:spr}, choose a
scaling range by eye estimation (this range is the same is all the snapshots)
and fit the log-log plot of the data
with a linear function to obtain the equal time scaling exponents.
In the worst case we obtain close to a decade of clean scaling. 
The structure functions from one such snapshot is shown in \Fig{fig:eqstr}.
Thus we obtain $M$ independent estimates for each exponent.
The mean of these $M$ values is our estimates of the equal time exponets
and the standard deviation is the error.

For the scaling exponents obtained from the exit times,
we introduce $2n$ tracer pairs, where $n=4096^2$,
in the statistically stationary state
and select only thoese whose  initial and their final separation
are within the inertial range.
Thence we calculate the negative moment of the exit times
and plot them, in log-log scale, as a function of $R$
and use linear fit to estimate the dynamic scaling exponents.
We repeat this exercise for $M=20$ independent initial
realization of the tracer pairs. 
For each exponent, its final estimate is its ensemble mean, and
its error bars are defined by the standard deviation.
These are the values plotted in \Fig{fig:chi}.

This method of estimating parameters of a model and its errors
from an ensemble based approach is often called BAGGing or Bootstrap
AGGregation method, see e.g., Ref.~\citep{mehta2019high} section 8.2.

\end{document}